\documentclass[english,pra,reprint]{revtex4-1}

\usepackage{mathpazo}
\usepackage[T1]{fontenc}
\usepackage[latin9]{inputenc}
\usepackage{babel}
\usepackage{amsthm}
\usepackage{amsmath}
\usepackage{amssymb}
\usepackage{graphicx}
\usepackage{esint}
\usepackage[unicode=true,pdfusetitle,
 bookmarks=true,bookmarksnumbered=false,bookmarksopen=false,
 breaklinks=true,pdfborder={0 0 0},backref=false,colorlinks=false]
 {hyperref}

\makeatletter

\allowdisplaybreaks

\DeclareMathOperator{\tr}{tr}

\makeatother

\begin{document}

\title{Polarimetry by classical ghost diffraction }

\author{Henri Kellock$^{1}$}
\email{henri.kellock@uef.fi}
\author{Tero Setälä$^{1}$}
\author{Ari T.~Friberg$^{1}$}
\author{Tomohiro Shirai$^{2}$}
\address{$^{1}$Institute of Photonics, University of Eastern Finland, P.O.
Box 111, FI-80101 Joensuu, Finland \\
$^{2}$Research Institute of Instrumentation Frontier, National Institute
of Advanced Industrial Science and Technology, AIST Tsukuba Central
2, 1-1-1 Umezono, Tsukuba 305-8568, Japan}

\begin{abstract}

We present a technique for studying the polarimetric properties of
a birefringent  object by means of classical ghost diffraction. The standard
ghost diffraction setup is modified to include polarizers for controlling
the state of polarization of the beam in various places. The object
is characterized by a Jones matrix and the absolute values of the
Fourier transforms of its individual elements are measured. From these
measurements the original complex-valued functions can be retrieved through
iterative methods resulting in the full Jones matrix of the object.
We present two different placements of the polarizers and show that
one of them leads to better polarimetric quality, while the other
placement  offers the possibility to perform polarimetry without
controlling the source's state of polarization. The concept of an
effective source is introduced to simplify the calculations. Ghost
polarimetry enables the assessment of polarization properties as a
function of position within the object through simple intensity correlation
measurements.

\end{abstract}
\maketitle

\section{Introduction}

Ghost imaging, or correlation imaging, is an imaging technique where
quantum entanglement and photon coincidences or classical intensity
correlation statistics are used to provide information on an object
through indirect measurements \cite{GATTI08,Erkmen10}.   Setups
can be arranged to form either the image or the far-field diffraction
pattern of the object.  The classical counterpart of ghost imaging
has its merits in readily-available and cost-effective, bright light
sources. It is able to emulate most aspects of quantum ghost imaging
\cite{Gatti04a,Gatti04b,Valencia05,Ferri05,Gatti06,Torres08}, although
classical correlation imaging has limitations with regard to the visibility
(contrast) of the resulting image \cite{Cao08,Chan10,Kellock12a}.
Techniques have been developed to overcome these limitations, including
higher-order correlations \cite{Cao08,Kellock12a}, background subtraction
\cite{Chan10}, differential ghost imaging \cite{Ferri10} and computational
ghost imaging \cite{Shapiro08,Bromberg09}. Using electromagnetic
theory also the effects that the degree of polarization of the light
has on correlation imaging have been studied recently \cite{Liu10,Tong10,Shirai11a,Kellock12a}.

In this paper we introduce a method for obtaining the polarimetric
properties of a (partially) transparent, spatially dependent, arbitrarily
birefringent or dichroic object by means of correlation imaging.
In contrast to 
quantum ellipsometry with two-photon polarization-entangled
light in reflection 
geometry 
reported previously \cite{Abouraddy01b,Abouraddy02b,Toussaint04},
we consider light from a classical, spatially incoherent, partially polarized
source interacting with the object in transmission arrangement.
We make use of electromagnetic theory of optical coherence and 
employ a modified ghost diffraction setup, with the concept of
an effective source used for convenience of analysis.  Whereas elaborate
techniques for measuring an object's polarization properties based
on Mueller imaging and Stokes imaging have been developed \cite{Brosseau1985},
in many applications (such as thin-film optics and nanophotonics)
a simpler polarimetric characterization of the object by means of
its Jones matrix is adequate. The conventional way of measuring the
Jones matrix of a spatially uniform, planar object makes use of a
coherent laser, two polarizers and a detector \cite{Brosseau1985}.
In correlation imaging light from a spatially incoherent source is
split into two arms, a test arm containing the object followed by
a non-resolving detector and a reference arm with a CCD camera. The
intensities recorded at the end of the two arms are correlated to
obtain either the image or the far-field diffraction pattern of the
object.  For polarimetry by ghost diffraction we present two possible
arrangements. In one, polarizers are  placed after the source and
before the detectors.  In the other arrangement the source's state
of polarization is not controlled and the polarizers are located in
front of the detectors in each arm. In both cases measurements are
done with different orientations of the polarizers to individually
select  all the elements of the object's Jones matrix.  The advantage
of ghost polarimetry is that spatially varying, polarization-state
altering objects can be analyzed. 

Section \ref{sec:Jones-matrix-of} recalls briefly the Jones calculus
and  the conventional measurement of the Jones matrix components.
In section \ref{sec:Ghostdiffraction} we introduce the mathematical
notion of the effective source and incorporate it in our discussion
on electromagnetic ghost diffraction. In section \ref{sec:Ghost-polarimetry} the standard
ghost diffraction setup is modified in two different ways  so as
to enable the measurement of the polarimetric properties of an unknown
object. The  polarimetric qualities of these two  setups are assessed
in terms of image visibility. The main conclusions are summarized
in section \ref{sec:Conclusions}. 
Appendix \ref{sec:Matrix-calculus}
 gives further details on the possible placements of the polarizers
 and on the constraints the placements put upon the source.

\section{Jones formalism for polarimetry\label{sec:Jones-matrix-of}}

We employ spectral electromagnetic coherence theory to analyze polarimetry
by ghost diffraction. A realization of a random, beam-like electric
field at position $\mathbf{r}$ and frequency $\omega$, propagating
in the $z$ direction is denoted by  $\mathbf{E}(\mathbf{r})=[E_{x}(\mathbf{r})\, E_{y}(\mathbf{r})]^{\mathrm{T}}$,
where the superscript $\mathrm{T}$ denotes the transpose and the
frequency dependence is suppressed for brevity. On illuminating with a linear, polarimetric
(birefringent, dichroic), planar optical element $\mathbf{T}(\mathbf{r})$
the output electric field $\mathbf{E}_{\mathrm{out}}(\mathbf{r})$
is related to the input electric field $\mathbf{E}_{\mathrm{in}}(\mathbf{r})$
through  \cite{Saleh2007} 
\begin{alignat}{1}
\mathbf{E}_{\mathrm{out}}(\mathbf{r}) & =\mathbf{T}(\mathbf{r})\mathbf{E}_{\mathrm{in}}(\mathbf{r}),\label{eq:et01}
\end{alignat}
 where 
\begin{alignat}{1}
\mathbf{T}(\mathbf{r}) & =\begin{bmatrix}T_{xx}(\mathbf{r}) & T_{xy}(\mathbf{r})\\
T_{yx}(\mathbf{r}) & T_{yy}(\mathbf{r})
\end{bmatrix}\label{eq:jones01}
\end{alignat}
 is the optical element's transmission (or Jones) matrix. We take
$T_{ij}\in\mathbb{C}$ with $i,j\in\{x,y\}$ to be deterministic and,
in general, spatially dependent when $\mathbf{T}$ represents the
object. For polarizers (and wave plates) the elements of $\mathbf{T}$
are constant. For example, 
\begin{alignat}{2}
\mathbf{T}^{x} & =\begin{bmatrix}1 & 0\\
0 & 0
\end{bmatrix}, & \,\,\,\,\,\,\mathbf{T}^{y} & =\begin{bmatrix}0 & 0\\
0 & 1
\end{bmatrix}\label{eq:tx01}
\end{alignat}
 are uniform linear polarizers (LPs) that only let the $x$ or $y$
component of the light  go through, unaltered.

\begin{figure}
\begin{centering}
\includegraphics[width=0.6\columnwidth]{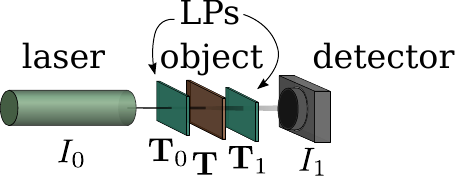}
\par\end{centering}
\caption{{\small  Measuring the elements of the matrix $\mathbf{T}$
that describes the response of the object. The setup consists of a
spatially coherent laser source ($I_{0}$), the object which is placed
between two linear polarizers (LPs), $\mathbf{T}_{0}$ and $\mathbf{T}_{1}$,
and a detector. The measured intensity ($I_{1}$) is compared to the
case where the object and the second LP are removed.  }\label{fig:polarimetry} }
\end{figure}

The conventional way to determine the polarimetric response of an unknown optical
 element is to measure the components of the corresponding Jones
matrix $\mathbf{T}$ individually. Normally, for a uniform object,
this is done with a setup schematically depicted in figure \ref{fig:polarimetry}.
A spatially coherent, uniformly polarized, monochromatic laser beam
illuminates the object which is sandwiched between two LPs, the `polarizer'
$\mathbf{T}_{0}$ and the `analyzer' $\mathbf{T}_{1}$ and the transmitted
intensity is recorded with a  detector. To avoid effects of diffraction,
the elements $\mathbf{T}_{0}$ and $\mathbf{T}_{1}$ and the detector
are deep in the Fresnel zone of the object $\mathbf{T}$.  According
to equation (\ref{eq:et01}) the field $\mathbf{E}_{1}$ at the detector
is described by 
\begin{alignat}{1}
\mathbf{E}_{1} & =\mathbf{T}_{1}\mathbf{T}\mathbf{T}_{0}\mathbf{E}_{0},\label{eq:e2t01}
\end{alignat}
 where $\mathbf{E}_{0}$ is the coherent electric field of the laser
source.  Orientating the LPs so that $\mathbf{T}_{0}=\mathbf{T}^{i}$
and $\mathbf{T}_{1}=\mathbf{T}^{j}$, with $i,j\in\{x,y\}$, we obtain,
from equations (\ref{eq:tx01}) and (\ref{eq:e2t01}), the intensity 
\begin{alignat}{1}
I_{1} & =\mathbf{E}_{1}^{\dagger}\mathbf{E}_{1}=\left|T_{ji}\right|^{2}\left|E_{0,i}\right|^{2},\label{eq:e2t02}
\end{alignat}
 where the dagger denotes the Hermitian conjugate and $|E_{0,i}|^{2}=I_{0,i}$
are the intensities of the source components. These can be measured
by removing $\mathbf{T}$ and $\mathbf{T}_{1}$ from the setup and
varying $\mathbf{T}_{0}$. Hence we can calculate $|T_{ji}|$ with
all $i,j\in\{x,y\}$.

Equation (\ref{eq:e2t02}) demonstrates that only the absolute values
$|T_{ij}|$ of the object's Jones matrix are obtained with this method.
Using a considerably more complicated system one could measure all
the complex transmission matrix parameters $T_{ij}$ of a spatially
independent object. Such a setup involves linear polarizers oriented
at $\pm45$ degrees with respect to the $x$ axis as well as circular
polarizers in addition to the LPs introduced earlier \cite{Brosseau1985}.
If the object has spatial dependence, the detector in figure \ref{fig:polarimetry}
may in principle be replaced with a CCD array.

\section{Ghost diffraction\label{sec:Ghostdiffraction}}

After more than a decade of  research, ghost imaging and diffraction
are now well-established optical techniques \cite{GATTI08,Erkmen10}.
In particular, lensless ghost diffraction is employed to obtain the
far-field pattern, or the Fourier transform, of the object \cite{Cai04,Cheng04,Zhang07}.
Classical ghost imaging and diffraction are conveniently analyzed
using optical coherence theory \cite{Shirai11b,Shirai12a}. In what
follows we introduce also the novel idea of an `effective source'
which  greatly simplifies the calculations.

\subsection{Intensity and field correlations}

We need to compute the correlations between the intensities in the
two arms of the ghost diffraction setup, i.e., the intensities $I_{\alpha}=\mathbf{E}_{\alpha}^{\dagger}\mathbf{E}_{\alpha}$
with $\alpha\in\{1,2\}$. Besides the frequency $\omega$, in this
section we further suppress the spatial dependencies of the functions
for notational brevity. We begin by dividing $I_{\alpha}$ into the
intensities of the $x$ and $y$ components of the electric field
as $I_{\alpha}=I_{\alpha,x}+I_{\alpha,y}$. The intensity correlation
can then be expressed as  
\begin{alignat}{1}
\left\langle I_{1}I_{2}\right\rangle  & =\sum_{i,j\in\{x,y\}}\left\langle I_{1,i}I_{2,j}\right\rangle ,\label{eq:ii01}
\end{alignat}
 where $\left\langle \ldots\right\rangle $ denotes the ensemble
average. Assuming that the field fluctuations of the source obey Gaussian
statistics, we may use the Gaussian moment theorem to write $\langle I_{1,i}I_{2,j}\rangle$
in terms  of second-order field correlations as \cite{Mandel1995,Tong10,Liu10,Kellock12a}
\begin{alignat}{1}
\left\langle I_{1,i}I_{2,j}\right\rangle  & =W_{11,ii}W_{22,jj}+\left|W_{12,ij}\right|^{2},\label{eq:ii02}
\end{alignat}
 where $W_{\alpha\beta,ij}=\langle E_{\alpha,i}^{*}E_{\beta,j}\rangle$,
$\alpha,\beta\in\{1,2\}$, $i,j\in\{x,y\}$, is the cross-spectral
density function and $E_{\alpha,i}$ denotes the $i$ component of
the field in arm $\alpha$.

Combining equations (\ref{eq:ii01}) and (\ref{eq:ii02}) and making
use of the cross-spectral density matrix (CSDM) 
\begin{alignat}{1}
\mathbf{W}_{\alpha\beta} & =\begin{bmatrix}W_{\alpha\beta,xx} & W_{\alpha\beta,xy}\\
W_{\alpha\beta,yx} & W_{\alpha\beta,yy}
\end{bmatrix},
\end{alignat}
we can write the intensity correlation in a compact form as \cite{Shirai11a,Kellock12a}
\begin{alignat}{1}
\left\langle I_{1}I_{2}\right\rangle  & =\tr\mathbf{W}_{11}\tr\mathbf{W}_{22}+\tr\mathbf{W}_{12}^{\dagger}\mathbf{W}_{12},\label{eq:i1i2}
\end{alignat}
 where $\tr$ denotes the trace, $\tr\mathbf{W}_{\alpha\alpha}=\langle I_{\alpha}\rangle$
is the average intensity  in the $\alpha$th arm and $\tr\mathbf{W}_{12}^{\dagger}\mathbf{W}_{12}=\langle\Delta I_{1}\Delta I_{2}\rangle$
 is the correlation of the intensity fluctuations $\Delta I_{\alpha}\equiv I_{\alpha}-\langle I_{\alpha}\rangle$
between the two different arms.

\subsection{Effective source\label{sec:Effective-source}}

\begin{figure}
\includegraphics[width=0.9\columnwidth]{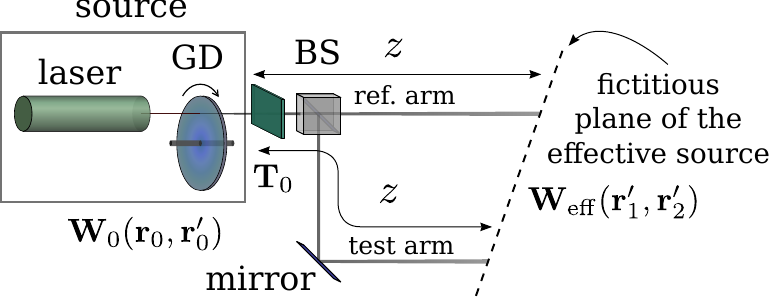}
\caption{{\small Concept of the effective source. A laser and
a rotating ground glass disk (GD) create a spatially incoherent planar
field. The polarizer $\mathbf{T}_{0}$, controlling the state of polarization
of the source, may or may not exist before the beam splitter (BS)
which directs the light into the reference arm and the test arm. The
beams propagate over the distance $z$ in both arms until reaching
the fictitious plane of the effective source. }\label{fig:effsource} }
\end{figure}

Before proceeding into the technical details of ghost diffraction,
we present with reference to figure \ref{fig:effsource} the notion
of the effective source. Coherent laser radiation passed through a
rotating disk of ground glass generates a stationary, uniformly polarized,
spatially incoherent beam of light characterized by the cross-spectral
density matrix $\mathbf{W}_{0}(\mathbf{r}_{0},\mathbf{r}_{0}^{\prime})$.
We then consider two possibilities: behind the physical source there
may be a linear polarizer represented by the matrix $\mathbf{T}_{0}$,
or it may be absent. A beam splitter separates the field into the
reference and test paths, followed by propagation over an equal distance
$z$ in each arm. The light characterized by the CSDM $\mathbf{W}_{\mathrm{eff}}(\mathbf{r}_{1}^{\prime},\mathbf{r}_{2}^{\prime})$
in the ensuing fictitious plane constitutes the effective source.

Within the accuracy of Fresnel diffraction, the effective source's
CSDM is given by (see figure \ref{fig:effsource})
\begin{alignat}{1}
\mathbf{W}_{\mathrm{eff}}(\mathbf{r}_{1}^{\prime},\mathbf{r}_{2}^{\prime}) & =\iint_{-\infty}^{\infty}\mathrm{d}^{2}r_{0}\mathrm{d}^{2}r_{0}^{\prime}\mathbf{W}(\mathbf{r}_{0},\mathbf{r}_{0}^{\prime})K_{1}^{*}(\mathbf{r}_{0},\mathbf{r}_{1}^{\prime})K_{2}(\mathbf{r}_{0}^{\prime},\mathbf{r}_{2}^{\prime})\label{eq:weff01}
\end{alignat}
where $\mathbf{W}(\mathbf{r}_{0},\mathbf{r}_{0}^{\prime})=\mathbf{W}_{0}(\mathbf{r}_{0},\mathbf{r}_{0}^{\prime})$
in the absence of the polarizer $\mathbf{T}_{0}$ and $\mathbf{W}(\mathbf{r}_{0},\mathbf{r}_{0}^{\prime})=\mathbf{T}_{0}^{*}\mathbf{W}_{0}(\mathbf{r}_{0},\mathbf{r}_{0}^{\prime})\mathbf{T}_{0}^{\mathrm{T}}$
in the presence of $\mathbf{T}_{0}$. Further, the kernel \cite{Goodman2005,Shirai11b}
\begin{alignat}{1}
K_{\alpha}(\mathbf{r}_{0},\mathbf{r}_{\alpha}^{\prime}) & =\frac{-ik}{2\pi z}\exp\left[\frac{ik}{2z}\left(\mathbf{r}_{\alpha}^{\prime}-\mathbf{r}_{0}\right)^{2}\right],\label{eq:k01}
\end{alignat}
with $\alpha\in\{1,2\}$ and $k=\omega/c$ being the wave number ($c$
is the vacuum speed of light), describes field  propagation over
the distance $z$ in free space from the physical source plane to
the fictitious plane of the effective source.

Now, in classical ghost diffraction the actual source is spatially
completely incoherent. Hence we may take $\mathbf{W}_{0}(\mathbf{r}_{0},\mathbf{r}_{0}^{\prime})=\mathbf{J}_{0}\delta(\mathbf{r}_{0}^{\prime}-\mathbf{r}_{0})$,
where $\mathbf{J}_{0}$ is the polarization matrix of the source and
$\delta\left(\mathbf{r}\right)$ is the two-dimensional Dirac delta
function. Applying the $\delta$ function to eliminate $\mathbf{r}_{0}^{\prime}$
  and using equation (\ref{eq:k01}), equation (\ref{eq:weff01}) then becomes
\begin{alignat}{1}
\mathbf{W}_{\mathrm{eff}}(\mathbf{r}_{1}^{\prime},\mathbf{r}_{2}^{\prime}) & =\mathbf{J}_{\mathrm{eff}}\delta(\mathbf{r}_{2}^{\prime}-\mathbf{r}_{1}^{\prime}),\label{eq:weff00p}
\end{alignat}
where $\mathbf{J}_{\mathrm{eff}}=\mathbf{T}_{0}^{*}\mathbf{J}_{0}\mathbf{T}_{0}^{\mathrm{T}}$
with $\mathbf{T}_{0}$ and $\mathbf{J}_{\mathrm{eff}}=\mathbf{J}_{0}$
without $\mathbf{T}_{0}$.   Equation (\ref{eq:weff00p}) shows
that in the case of a (sufficiently wide) spatially completely incoherent
physical source also the effective source is spatially incoherent.

\subsection{Fourier transform by ghost diffraction}

\begin{figure}[b]
\includegraphics[width=0.9\columnwidth]{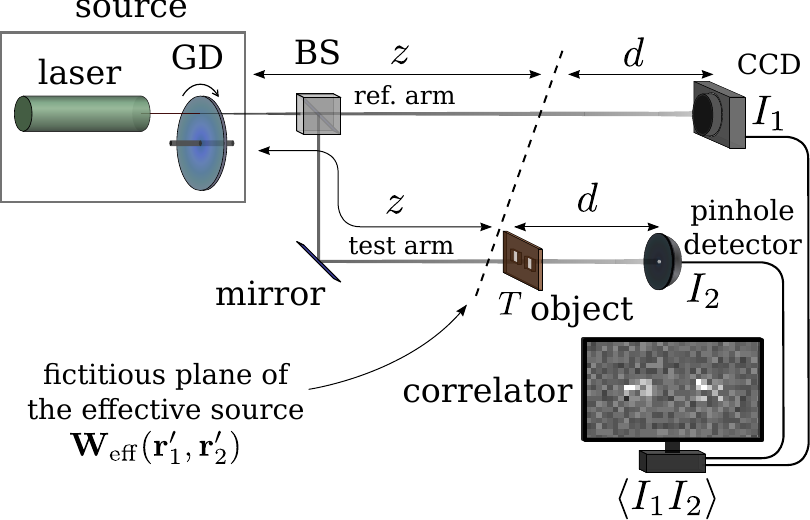}
\caption{{\small Effective source in ghost diffraction. The
fictitious plane is at a distance $z$ from the true source and contains
the effective source characterized by $\mathbf{W}_{\mathrm{eff}}(\mathbf{r}_{1}^{\prime},\mathbf{r}_{2}^{\prime})$
(see figure }\ref{fig:effsource}{\small ). In the reference arm the
wave propagates a further distance $d$ and is detected by a CCD camera.
In the test arm the object $T$ is located immediately after the effective
source and the field then travels the same distance $d$ onto a pinhole
detector. The measured intensities $I_{1}$ and $I_{2}$ are correlated.
   }\label{fig:ghostimaging} }
\end{figure}

We next demonstrate that in the context of classical ghost diffraction
the concept of the effective source  is consistent with the formation
of the Fourier transform of the object distribution. We represent
the object here as $\mathbf{T}(\mathbf{r}^{\prime})=T(\mathbf{r}^{\prime})\mathbf{I}$,
where $\mathbf{I}$ is the $2\times2$ unit matrix and omit the polarizer
$\mathbf{T}_{0}$ that controls the source's polarization state, as
is illustrated in figure \ref{fig:ghostimaging}. In the  test arm
the object is placed immediately after the effective source. The fields
then propagate from the effective source plane an equal distance $d$
in both arms to the detectors, which are taken to be pointlike.

Using an expression of the form of equation (\ref{eq:weff01}) with integrations
over the effective source, we obtain for the CSDM between the electric
fields in the two arms the formula
\begin{alignat}{1}
\mathbf{W}_{12}(\mathbf{r}_{1},\mathbf{r}_{2}) & =\iint_{-\infty}^{\infty}\mathrm{d}^{2}r_{1}^{\prime}\mathrm{d}^{2}r_{2}^{\prime}\mathbf{W}_{\mathrm{eff}}(\mathbf{r}_{1}^{\prime},\mathbf{r}_{2}^{\prime})\nonumber \\
 & \times T(\mathbf{r}_{2}^{\prime})K_{1}^{*}(\mathbf{r}_{1}^{\prime},\mathbf{r}_{1})K_{2}(\mathbf{r}_{2}^{\prime},\mathbf{r}_{2}),\label{eq:w01}
\end{alignat}
 where $T(\mathbf{r}_{2}^{\prime})$ is the  object's transmission
function. On further employing equations (\ref{eq:k01}) and (\ref{eq:weff00p})
with $z=d$ and $\mathbf{J}_{\mathrm{eff}}=\mathbf{J}_{0}$, respectively,
and integrating over the variable $\mathbf{r}_{2}^{\prime}$, equation (\ref{eq:w01})
becomes

\begin{alignat}{1}
\mathbf{W}_{12}(\mathbf{r}_{1},\mathbf{r}_{2}) & =\frac{k^{2}\mathbf{J}_{0}}{2\pi d^{2}}\exp\left[\frac{ik}{2d}\left(\mathbf{r}_{2}^{2}-\mathbf{r}_{1}^{2}\right)\right]\nonumber \\
 & \times\mathcal{F}\left\{ T(\mathbf{r}_{1}^{\prime})\right\} \left[\frac{k}{d}\left(\mathbf{r}_{1}-\mathbf{r}_{2}\right)\right],\label{eq:w03}
\end{alignat}
where  
\begin{alignat}{1}
\mathcal{F}\left\{ T(\mathbf{r})\right\} \left[\mathbf{k}\right] & =\frac{1}{2\pi}\int_{-\infty}^{\infty}\mathrm{d}^{2}rT(\mathbf{r})e^{i\mathbf{k}\cdot\mathbf{r}}\label{eq:fourier01}
\end{alignat}
 is the two-dimensional Fourier transform of $T(\mathbf{r})$.  Since
in the test path we employ a pinhole detector (see figure \ref{fig:ghostimaging}),
we may naturally take it to be located at the point $\mathbf{r}_{2}=\mathbf{0}$
and so equation (\ref{eq:w03}) reduces to 
\begin{alignat}{1}
\mathbf{W}_{12}(\mathbf{r}_{1},\mathbf{0}) & =\frac{k^{2}\mathbf{J}_{0}}{2\pi d^{2}}\exp\left(-\frac{ik}{2d}\mathbf{r}_{1}^{2}\right)\mathcal{F}\left\{ T(\mathbf{r}_{1}^{\prime})\right\} \left[\frac{k}{d}\mathbf{r}_{1}\right].\label{eq:w04}
\end{alignat}
 Hence, using a  spatially uncorrelated light source and the appropriate
path-length conditions for classical ghost diffraction ($z+d$ in
both arms),  we have obtained the Fourier transform of the object
via the notion of the effective source.  We will next apply these
same ideas to ghost polarimetry.

\section{Ghost polarimetry\label{sec:Ghost-polarimetry}}

Making use of the effective source concept we present two different
modifications to the classical ghost diffraction setup which 
are used to obtain the moduli of the Fourier transforms of each of
the elements in the object's Jones matrix. Applying standard iterative
methods all elements of the complex Jones matrix can then, in principle,
be computed \cite{Shirai12a,McBride04}. We further assess separately
the polarimetric image qualities of the two arrangements. One of the
modifications leads to a setup which can be employed to perform polarimetry
on an arbitrary birefringent or dichroic object without a polarization-state
controlled source, whereas the other gives a better image contrast
in a configuration that closer resembles the conventional polarimetric
setup.

\subsection{Configurations for ghost polarimetry\label{sub:Ghost-polarimetry}}

\begin{figure}
\includegraphics[width=0.9\columnwidth]{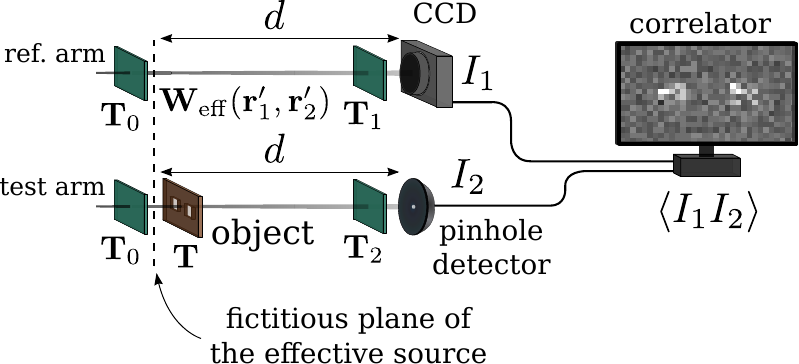}
\caption{{\small Ghost polarimetry configurations. The first
polarizer $\mathbf{T}_{0}$ is placed before the plane of the effective
source $\mathbf{W}_{\mathrm{eff}}(\mathbf{r}_{1}^{\prime},\mathbf{r}_{2}^{\prime})$
and affects both arms. In the reference arm another polarizer $\mathbf{T}_{1}$
is placed before the CCD camera. In the test arm the object $\mathbf{T}$
is located immediately after the effective source and another polarizer
$\mathbf{T}_{2}$ is placed before the pinhole detector. 
The intensities $I_{1}$ and $I_{2}$ are measured at the end of the arms and correlated. 
}\label{fig:ghostpolarimetry} }
\end{figure}

For ghost polarimetry of an arbitrary planar object we modify the
setup shown in figure \ref{fig:ghostimaging} slightly. More specifically,
we employ the effective source with the  possibility of controlling
the actual source's state of polarization with the uniform polarizer
$\mathbf{T}_{0}$ (see figure \ref{fig:effsource}). In addition, spatially
independent linear polarizers described by $\mathbf{T}_{1}$ and $\mathbf{T}_{2}$
are placed  before the CCD camera in the reference arm and in front
of the pinhole detector in the test arm, respectively. The object's
matrix transmission function is  $\mathbf{T}(\mathbf{r}_{2}^{\prime})$,
explicitly given by equation (\ref{eq:jones01}) and it can alter the
state of polarization of the test beam. The simplified setup, without
a detailed description of the effective source, is shown in figure \ref{fig:ghostpolarimetry}.

In analogy with equation (\ref{eq:w01}) the CSDM between the light in
the two arms now becomes {[}see also equation (\ref{eq:et01}){]} 
\begin{alignat}{1}
\mathbf{W}_{12}(\mathbf{r}_{1},\mathbf{r}_{2}) & =\iint_{-\infty}^{\infty}\mathrm{d}^{2}r_{1}^{\prime}\mathrm{d}^{2}r_{2}^{\prime}\mathbf{T}_{1}^{*}\mathbf{W}_{\mathrm{eff}}(\mathbf{r}_{1}^{\prime},\mathbf{r}_{2}^{\prime})\mathbf{T}^{\mathrm{T}}(\mathbf{r}_{2}^{\prime})\mathbf{T}_{2}^{\mathrm{T}}\nonumber \\
 & \times K_{1}^{*}(\mathbf{r}_{1}^{\prime},\mathbf{r}_{1})K_{2}(\mathbf{r}_{2}^{\prime},\mathbf{r}_{2}),\label{eq:w01p}
\end{alignat}
   where the integrations are in the effective source plane and
the propagation kernels $K_{\alpha}(\mathbf{r}_{\alpha}^{\prime},\mathbf{r}_{\alpha})$,
with and $\alpha\in\{1,2\}$, are given by equation (\ref{eq:k01}) with
$z=d$. On inserting the CSDM of the effective source from equation (\ref{eq:weff00p})
and the explicit forms of $K_{\alpha}(\mathbf{r}_{\alpha}^{\prime},\mathbf{r}_{\alpha})$,
 and integrating with respect to $\mathbf{r}_{2}^{\prime}$, we obtain
\begin{alignat}{1}
\mathbf{W}_{12}(\mathbf{r}_{1},\mathbf{r}_{2}) & =\frac{k^{2}}{4\pi^{2}d^{2}}\int_{-\infty}^{\infty}\mathrm{d}^{2}r_{1}^{\prime}\mathbf{T}_{1}^{*}\mathbf{J}_{\mathrm{eff}}\mathbf{T}^{\mathrm{T}}(\mathbf{r}_{1}^{\prime})\mathbf{T}_{2}^{\mathrm{T}}\nonumber \\
 & \times\exp\left\{ \frac{ik}{2d}\left[\mathbf{r}_{2}^{2}-\mathbf{r}_{1}^{2}-2\left(\mathbf{r}_{2}-\mathbf{r}_{1}\right)\cdot\mathbf{r}_{1}^{\prime}\right]\right\} .\label{eq:w02p}
\end{alignat}
To perform the polarimetric measurements, the polarizers are oriented
in such a manner that the conditions $\mathbf{T}_{1}^{*}\mathbf{J}_{\mathrm{eff}}=J_{0,ii}\mathbf{T}^{i}$
and $\mathbf{T}_{2}^{\mathrm{T}}=\mathbf{T}^{j}$ are met, with $i,j\in\{x,y\}$
and $J_{0,ii}$ denoting a diagonal component of $\mathbf{J}_{0}$.
Specific arrangements that result in the former requirement are discussed
in section \ref{sub:Image-quality}.  Using the aforementioned conditions
together with  equations (\ref{eq:tx01})  and (\ref{eq:w02p}), we
find that {[}compare with equation (\ref{eq:w03}){]} 
\begin{alignat}{1}
\mathbf{W}_{12}(\mathbf{r}_{1},\mathbf{r}_{2}) & =\frac{k^{2}J_{0,ii}\mathbf{T}^{ij}}{2\pi d^{2}}\exp\left[\frac{ik}{2d}\left(\mathbf{r}_{2}^{2}-\mathbf{r}_{1}^{2}\right)\right]\nonumber \\
 & \times\mathcal{F}\left\{ T_{ji}(\mathbf{r}_{1}^{\prime})\right\} \left[\frac{k}{d}\left(\mathbf{r}_{1}-\mathbf{r}_{2}\right)\right],\label{eq:w03p}
\end{alignat}
 where $\mathcal{F}\{T_{ji}(\mathbf{r}_{1}^{\prime})\}[k(\mathbf{r}_{1}-\mathbf{r}_{2})/d]$
is the Fourier transform of $T_{ji}(\mathbf{r}_{1}^{\prime})$, as
defined by equation (\ref{eq:fourier01}) and the matrix $\mathbf{T}^{ij}$
has the element $T_{ij}^{ij}=1$, with all the other elements being
zero.  With reference to figure \ref{fig:ghostpolarimetry} we again
invoke the condition that the pinhole detector in the test arm is
placed at the position $\mathbf{r}_{2}=\mathbf{0}$. We then find,
in analogy with equation (\ref{eq:w04}), that 
\begin{alignat}{1}
\mathbf{W}_{12}(\mathbf{r}_{1},\mathbf{0}) & =\frac{k^{2}J_{0,ii}\mathbf{T}^{ij}}{2\pi d^{2}}\exp\left(-\frac{ik}{2d}\mathbf{r}_{1}^{2}\right)\mathcal{F}\left\{ T_{ji}(\mathbf{r}_{1}^{\prime})\right\} \left[\frac{k}{d}\mathbf{r}_{1}\right].\label{eq:w04p}
\end{alignat}
Consequently, the  correlation of the intensity fluctuations between
the two arms is 
\begin{alignat}{1}
\tr\mathbf{W}_{12}^{\dagger}\mathbf{W}_{12} & =\frac{k^{4}J_{0,ii}^{2}}{4\pi^{2}d^{4}}\left|\mathcal{F}\left\{ T_{ji}(\mathbf{r}_{1}^{\prime})\right\} \left[\frac{k}{d}\mathbf{r}_{1}\right]\right|^{2}.\label{eq:trw04p}
\end{alignat}
 Note that we must separately measure the intensity fluctuation correlation
for all $i,j\in\{x,y\}$. 

Once the moduli of the Fourier transforms are known, the complex elements
can be calculated using iterative methods \cite{Shirai12a,McBride04},
and effects like linear birefringence and linear or circular dichroism
can be observed. For scalar amplitude-only and pure phase objects
\cite{Elser03} the iterative techniques have been demonstrated in
the realm of ghost imaging \cite{Zhang07}. Phase-contrast ghost imaging
could be used for retrieval of phase-only objects \cite{Shirai11b}.

\subsection{ Polarimetric qualities in the two setups\label{sub:Image-quality}}

To obtain equation (\ref{eq:trw04p}) we assumed that $\mathbf{T}_{1}^{*}\mathbf{J}_{\mathrm{eff}}=J_{0,ii}\mathbf{T}^{i}$.
This condition holds, for instance, when the polarizer in the reference
arm is disregarded ($\mathbf{T}_{1}=\mathbf{I}$), the polarizer after
the source satisfies $\mathbf{T}_{0}=\mathbf{T}^{i}$ and the actual
source is polarized at an angle of 45 degrees with respect to the
$x$ axis. We call this setup case A. Another arrangement in which
the condition $\mathbf{T}_{1}^{*}\mathbf{J}_{\mathrm{eff}}=J_{0,ii}\mathbf{T}^{i}$
holds is when the polarizer following the source  is discarded
resulting in $\mathbf{J}_{\mathrm{eff}}=\mathbf{J}_{0}$,  the polarizer
in the reference arm satisfies $\mathbf{T}_{1}=\mathbf{T}^{i}$, and
the polarization matrix $\mathbf{J}_{0}$ of the actual source is
diagonal as is true, e.g., for completely unpolarized light. This
setup, which we label case B, has the important property that it would
permit polarimetry without a polarization-state controlled source.
We recall that in both cases A and B the polarizer in the test arm
must satisfy $\mathbf{T}_{2}^{\mathrm{T}}=\mathbf{T}^{j}$. 

Although the main result {[}equation (\ref{eq:trw04p}){]} is the same
in  cases A and B, the polarimetric image qualities are different.
There are several methods for evaluating the image quality in ghost
imaging and diffraction, including multiple  definitions for the
visibility \cite{GATTI08,Cao08,Shirai11a}, the signal-to-noise ratio
(SNR) \cite{Ferri10,Brida11} and the contrast-to-noise ratio (CNR)
\cite{Chan10,Kellock12a}.  For a comparison between the cases A
and B  we assess the polarimetric quality in terms of the visibility
defined as \cite{Shirai11a} 
\begin{alignat}{1}
\mathrm{V} & \equiv\frac{\left\langle I_{1}I_{2}\right\rangle _{\mathrm{max}}-\left\langle I_{1}I_{2}\right\rangle _{\mathrm{min}}}{\left\langle I_{1}I_{2}\right\rangle _{\mathrm{max}}+\left\langle I_{1}I_{2}\right\rangle _{\mathrm{min}}},\label{eq:vis01}
\end{alignat}
 where the subscript $\mathrm{max}$ denotes the average over the
bright area of the image and the subscript $\mathrm{min}$ stands
for the average over the dark area. Using equation (\ref{eq:i1i2}) we
obtain 
\begin{alignat}{1}
\mathrm{V} & =\frac{[\tr\mathbf{W}_{12}^{\dagger}\mathbf{W}_{12}]_{\mathrm{max}}-[\tr\mathbf{W}_{12}^{\dagger}\mathbf{W}_{12}]_{\mathrm{min}}}{2\tr\mathbf{W}_{11}\tr\mathbf{W}_{22}+[\tr\mathbf{W}_{12}^{\dagger}\mathbf{W}_{12}]_{\mathrm{max}}+[\tr\mathbf{W}_{12}^{\dagger}\mathbf{W}_{12}]_{\mathrm{min}}}.\label{eq:vis02}
\end{alignat}
 The visibility thus depends not only on the correlation of the intensity
fluctuations between the reference and test arms ($\tr\mathbf{W}_{12}^{\dagger}\mathbf{W}_{12}$)
but also on the background term in ghost imaging, i.e., the product
of the intensities measured in the reference ($\tr\mathbf{W}_{11}$)
and test ($\tr\mathbf{W}_{22}$) arms. As already noted,  the CSDM
$\mathbf{W}_{12}$ between the light in the reference and test arms
is the same in the two cases A and B, since  each case leads to equation (\ref{eq:w04p})
and subsequently equation (\ref{eq:trw04p}) is obtained. 

Also the intensities in the reference arm are identical in both cases
A and B, as moving the polarizer from  the vicinity of the source
to in front of the reference arm detector does not change anything
for that arm. The explicit form of the CSDM for the reference arm
is calculated in 
appendix \ref{sec:Matrix-calculus},
equation (\ref{eq:w04p11}).
Its trace then is
\begin{alignat}{1}
\tr\mathbf{W}_{11} & =J_{0,ii}.\label{eq:trw04p11}
\end{alignat}
The average intensity in the reference arm is independent of transverse
position across the detector.

However, in the test arm the intensities are slightly different for
the two cases.  In case A the polarizer $\mathbf{T}_{0}$ placed
directly after the source can control the polarization state of the
light that goes into the reference arm detector as well as of the
light that traverses the object in the test arm. In this situation
we obtain, from equation (\ref{eq:w04p22}), 
\begin{alignat}{1}
\tr\mathbf{W}_{22} & =\frac{k^{2}J_{0,ii}}{4\pi^{2}d^{2}}\int_{-\infty}^{\infty}\mathrm{d}^{2}r_{1}^{\prime}\left|T_{ji}(\mathbf{r}_{1}^{\prime})\right|^{2}.\label{eq:trw04p22}
\end{alignat}
 On the other hand, in case B, in which the source is fixed and the
polarizer is in front of the reference arm detector, the light interacting
with the object has not been filtered from unnecessary polarization
states. From equation (\ref{eq:w04p22b}) we now have

\begin{alignat}{1}
\tr\mathbf{W}_{22} & =\frac{k^{2}}{4\pi^{2}d^{2}}\int_{-\infty}^{\infty}\mathrm{d}^{2}r_{1}^{\prime}\nonumber \\
 & \times\left(J_{0,xx}\left|T_{jx}(\mathbf{r}_{1}^{\prime})\right|^{2}+J_{0,yy}\left|T_{jy}(\mathbf{r}_{1}^{\prime})\right|^{2}\right),\label{eq:trw04p22b}
\end{alignat}
  The average intensity in the test arm {[}equations (\ref{eq:trw04p22})
and (\ref{eq:trw04p22b}){]} likewise is independent of transverse
position, albeit we need it only at the location of the pinhole detector.

The visibility $\mathrm{V}$ for case A can be calculated using equations (\ref{eq:trw04p})--(\ref{eq:trw04p22}).
For case B the last equation is replaced by the intensity given by
equation (\ref{eq:trw04p22b}). Comparing the two options  we note that
in case B  the intensity in the test arm is larger if a similar source
 is used and when all the object's transmission function's elements
are nonzero, i.e., $|T_{ij}(\mathbf{r}_{1}^{\prime})|>0$ for all
$i,j\in\{x,y\}$. From the form of equation (\ref{eq:vis02}) we may then
conclude that this results in lower visibility when compared to case
A. 
However, we emphasize that case B can be used in novel polarimetric
measurements, due to the fact that the classical source's state of polarization
does not need to be controlled.

\section{Conclusions\label{sec:Conclusions}}

By a modification to the conventional ghost diffraction experiment
with classical spatially incoherent light, a method for polarimetry
by ghost imaging has been introduced. Using the Jones matrix formalism
to analyze an electromagnetic ghost diffraction setup with polarization-dependent
optical elements, we have shown that adding two uniform linear polarizers
permits one to measure the Jones matrix of an arbitrary birefringent
or dichroic object. The first polarizer can be placed in front of
the reference arm detector to enable polarimetry without controlling
the state of polarization of the light source.   However, higher
polarimetric quality in terms of  image visibility is achieved when
the first polarizer is placed immediately after the source. The second
polarizer has to be located  between the object and the test arm detector
to discriminate how the object changes the different states of polarization
of the illumination. The general method presented in this work, called
ghost polarimetry, enables the assessment of the polarimetric properties
of spatially dependent objects in novel ways merely by means of intensity
correlations.

We have also demonstrated that the use of the effective source concept
greatly simplifies the calculations related to both scalar and electromagnetic
ghost diffraction setups. The effective source can be seen as a useful
mathematical tool for the analysis of a wide range of ghost imaging
and diffraction arrangements. It was shown that for sufficiently wide
beams the effective source retains the spatially completely incoherent
nature of the original source.

\section*{Acknowledgments}

Part of this work was done while HK was working with Aalto University.
HK acknowledges support from the Väisälä Fund granted by the Finnish
Academy of Science and Letters. This research was also supported by
the Academy of Finland (projects 272692, 268480 and 268705) and by the Japan Society for
the Promotion of Science (KAKENHI 23656054 and 25390104).

\appendix

\section{Placement of the optical elements\label{sec:Matrix-calculus}}

In ghost imaging, such as the setup depicted in figure \ref{fig:ghostpolarimetry},
the information is encoded in the correlation of the intensity fluctuations,
i.e., in $\tr\mathbf{W}_{12}^{\dagger}\mathbf{W}_{12}$. For the 
discussion in 
section \ref{sub:Necessary-polarizers}, 
the essential
part is the influence of the matrices $\mathbf{T}_{1}$, $\mathbf{J}_{\mathrm{eff}}$,
$\mathbf{T}(\mathbf{r}_{1}^{\prime})$ and $\mathbf{T}_{2}$  {[}see
equation (\ref{eq:w02p}){]} on  $\tr\mathbf{W}_{12}^{\dagger}\mathbf{W}_{12}$.
 We need to measure the  individual elements of $\mathbf{T}(\mathbf{r}_{1}^{\prime})$.
Using two different objects as an example, we show that leaving out
the polarizer between the object and the pinhole detector ($\mathbf{T}_{2}=\mathbf{I}$),
or leaving out both the polarizer after the source  ($\mathbf{J}_{\mathrm{eff}}=\mathbf{J}_{0}$)
and the polarizer before the CCD in the reference arm ($\mathbf{T}_{1}=\mathbf{I}$),
will make it impossible to distinguish between the objects.

In 
section \ref{sub:Example-measurements}
we present details
of the two cases in which the first polarizer is located in place
of either $\mathbf{T}_{0}$ or $\mathbf{T}_{1}$ and the second polarizer
is represented by $\mathbf{T}_{2}$. We examine each case separately
to find what kind of source is appropriate for the polarimetric imaging
and calculate the CSDMs for the individual arms. The CSDMs are used
in the discussion related to polarimetric quality in section \ref{sub:Image-quality}.

\subsection{Necessary polarizers\label{sub:Necessary-polarizers}}

With the goal of showing that leaving out certain polarizers  leads
to an unsuccessful measurement, first we consider the case in which
there is no polarizer after the source or in front of the reference
arm detector. In this case $\mathbf{J}_{\mathrm{eff}}=\mathbf{J}_{0}$
and $\mathbf{T}_{1}=\mathbf{I}$. The  polarization matrix  is Hermitian
and non-negative definite and is thus diagonalizable with a unitary
transformation $\mathbf{U}$ and non-negative eigenvalues $J_{1}$
and $J_{2}$. Let us perform the same transformation also to the other
matrices appearing in $\mathbf{W}_{12}$, i.e., \begin{subequations}
\begin{alignat}{1}
\mathbf{J}_{0}^{\prime} & =\mathbf{U}\mathbf{J}_{0}\mathbf{U}^{\dagger}=\begin{pmatrix}J_{1} & 0\\
0 & J_{2}
\end{pmatrix},\\
\mathbf{T}^{\prime\mathrm{T}}(\mathbf{r}_{1}^{\prime}) & =\mathbf{U}\mathbf{T}^{\mathrm{T}}(\mathbf{r}_{1}^{\prime})\mathbf{U}^{\dagger},\\
\mathbf{T}_{2}^{\prime\mathrm{T}} & =\mathbf{U}\mathbf{T}_{2}^{\mathrm{T}}\mathbf{U}^{\dagger}.
\end{alignat}
\label{eq:a234}\end{subequations}   We
present two example objects that can not be distinguished from each
other. The first one is represented by a diagonal matrix $\mathbf{T}^{\prime}(\mathbf{r}_{1}^{\prime})$
with the elements $T_{xx}^{\prime}(\mathbf{r}_{1}^{\prime})=g(\mathbf{r}_{1}^{\prime})$
and $T_{yy}^{\prime}(\mathbf{r}_{1}^{\prime})=\kappa g(\mathbf{r}_{1}^{\prime})$,
where $\kappa\equiv J_{1}/J_{2}$ and $g(\mathbf{r}_{1}^{\prime})$
is a Fourier-transformable complex function. The second object has
the non-zero elements $T_{xy}^{\prime}(\mathbf{r}_{1}^{\prime})=-\kappa g(\mathbf{r}_{1}^{\prime})$
and $T_{yx}^{\prime}(\mathbf{r}_{1}^{\prime})=g(\mathbf{r}_{1}^{\prime})$
(obtained by placing a $\pi/2$ rotator after the first object).
Combining  equations (\ref{eq:w02p}) and (\ref{eq:a234}) together with
the conditions $\mathbf{J}_{\mathrm{eff}}=\mathbf{J}_{0}$ and $\mathbf{T}_{1}=\mathbf{I}$,
we obtain
\begin{alignat}{1}
\tr\mathbf{W}_{12}^{\dagger}\mathbf{W}_{12} & =\frac{J_{1}^{2}k^{4}}{4\pi^{2}d^{4}}\left|\mathcal{F}\left\{ g(\mathbf{r}_{1}^{\prime})\right\} \left[\frac{k}{d}\left(\mathbf{r}_{1}-\mathbf{r}_{2}\right)\right]\right|^{2}\tr\mathbf{T}_{2}^{*}\mathbf{T}_{2}^{\mathrm{T}},\label{eq:trttt01}
\end{alignat}
 where $\mathcal{F}\{g(\mathbf{r}_{1}^{\prime})\}[k(\mathbf{r}_{1}-\mathbf{r}_{2})/d]$
is defined by equation (\ref{eq:fourier01}). This result was acquired
using the invariance of the trace under unitary transformations and
the intermediate result given by $\tr\mathbf{T}_{2}^{\prime*}\mathbf{T}^{\prime*}(\mathbf{r}_{1}^{\prime})\mathbf{J}_{0}^{\prime2}\mathbf{T}^{\prime\mathrm{T}}(\mathbf{r}_{1}^{\prime\prime})\mathbf{T}_{2}^{\prime\mathrm{T}}=J_{1}^{2}g^{*}(\mathbf{r}_{1}^{\prime})g(\mathbf{r}_{1}^{\prime\prime})\tr\mathbf{T}_{2}^{\prime*}\mathbf{T}_{2}^{\prime\mathrm{T}}$.
Equation (\ref{eq:trttt01}) holds for both example objects and we
can thus conclude that by varying $\mathbf{T}_{2}$ (or $\mathbf{T}_{2}^{\prime}$)
we are not able to make a distinction between the two cases, although
their Jones matrices are different.

Similar reasoning applies when the effective source and the polarizer
in front of the reference arm detector are variable but the test arm
detector lacks a polarizer ($\mathbf{T}_{2}=\mathbf{I}$). Using either
of the example objects presented above, in this situation we have
{[}see equation (\ref{eq:w02p}){]} 
\begin{alignat}{1}
\tr\mathbf{W}_{12}^{\dagger}\mathbf{W}_{12} & =\frac{k^{4}}{4\pi^{2}d^{4}}\left|\mathcal{F}\left\{ g(\mathbf{r}_{1}^{\prime})\right\} \left[\frac{k}{d}\left(\mathbf{r}_{1}-\mathbf{r}_{2}\right)\right]\right|^{2}\nonumber \\
 & \times\left(B_{xx}+\kappa^{2}B_{yy}\right),\label{eq:trttt02}
\end{alignat}
 where $B_{ii}$, $i\in\{x,y\}$, is the diagonal component of the
(generally non-diagonal) matrix $\mathbf{B}\equiv\mathbf{J}_{\mathrm{eff}}^{\dagger}\mathbf{T}_{1}^{\mathrm{T}}\mathbf{T}_{1}^{*}\mathbf{J}_{\mathrm{eff}}$
and we used $\tr\mathbf{T}^{*}(\mathbf{r}_{1}^{\prime})\mathbf{B}\mathbf{T}^{\mathrm{T}}(\mathbf{r}_{1}^{\prime\prime})=g^{*}(\mathbf{r}_{1}^{\prime})g(\mathbf{r}_{1}^{\prime\prime})(B_{xx}+\kappa^{2}B_{yy})$.
Since the result is the same for both objects, no matter how $\mathbf{J}_{\mathrm{eff}}$
and $\mathbf{T}_{1}$ are chosen, the objects are indistinguishable
when $\mathbf{T}_{2}=\mathbf{I}$.

In practice this means that ghost polarimetry  is not possible when
the polarizer between the object and the test arm detector is left
out ($\mathbf{T}_{2}=\mathbf{I}$), or if that optical element is
the only polarizer in the ghost imaging arrangement  ($\mathbf{J}_{\mathrm{eff}}=\mathbf{J}_{0}$
and $\mathbf{T}_{1}=\mathbf{I}$). This is analogous to the polarimetric
analysis performed by the conventional device shown in figure \ref{fig:polarimetry}
in the sense that leaving out the polarizer on either side of the
object would result in an unsuccessful measurement.

\subsection{Example measurements\label{sub:Example-measurements}}

In section \ref{sub:Image-quality} two arrangements for obtaining the
polarization properties of the object were introduced. As shown in
section \ref{sub:Ghost-polarimetry}, the intensity fluctuation correlation
($\tr\mathbf{W}_{12}^{\dagger}\mathbf{W}_{12}$) is the same in both
cases. However, the requirements on the source are slightly different
and are discussed in the following subsections. The CSDMs of the reference
($\mathbf{W}_{11}$) and test ($\mathbf{W}_{22}$) arms needed to
obtain the background term in equation (\ref{eq:vis02}) are evaluated.
They are used to assess the polarimetric quality in section \ref{sub:Image-quality}.

In the arrangement 
labeled 
case A,  the polarizer in front of the
reference arm detector is omitted ($\mathbf{T}_{1}=\mathbf{I}$) and
we choose $\mathbf{T}_{0}=\mathbf{T}^{i}$ for the polarizer after
the source and thus $\mathbf{J}_{\mathrm{eff}}=\mathbf{T}^{i}\mathbf{J}_{0}\mathbf{T}^{i}$.
In case B,  the polarizer is moved from after the source to\emph{
}in front of the reference arm detector and we have $\mathbf{J}_{\mathrm{eff}}=\mathbf{J}_{0}$
and $\mathbf{T}_{1}=\mathbf{T}^{i}$. In both cases the test arm has
the polarizer $\mathbf{T}_{2}=\mathbf{T}^{j}$.

\subsubsection{Case A: Polarizer after the source\label{sub:Optical-element-after}}

In this case, the effective source's polarization matrix, $\mathbf{J}_{\mathrm{eff}}=\mathbf{T}_{0}^{*}\mathbf{J}_{0}\mathbf{T}_{0}^{\mathrm{T}}$,
is proportional to $\mathbf{T}_{0}=\mathbf{T}^{i}$ when $\mathbf{J}_{0}$
is not polarized perpendicular to $\mathbf{T}^{i}$. This has to hold
for $i\in\{x,y\}$ and thus a source which is completely polarized
in either the $x$ or $y$ direction cannot be used. In order to obtain
the same intensity for both orientations ($i\in\{x,y\}$) of the polarizer
 $\mathbf{T}_{0}$,  the source should be either completely unpolarized,
(partially) circularly polarized, or (partially) linearly polarized
at $\pm45$ degrees from the $x$ axis.

 To calculate the CSDM related to the reference arm ($\mathbf{W}_{11}$),
we use a source slightly modified from equation (\ref{eq:weff00p}), with
the Dirac delta function replaced by a normalized coherence function
that has the property $\gamma(\mathbf{0})=1$, thus leading to the
CSDM $\mathbf{W}_{\mathrm{eff}}(\mathbf{r}_{1}^{\prime},\mathbf{r}_{2}^{\prime})=\mathbf{J}_{\mathrm{eff}}\gamma(\mathbf{r}_{2}^{\prime}-\mathbf{r}_{1}^{\prime})$.
Using a different source here is permitted for the qualitative visibility
comparison between the two cases we perform in section \ref{sub:Image-quality}.
The $\gamma$ source together with equations (\ref{eq:et01}), (\ref{eq:weff01}),
and (\ref{eq:k01})  produces the reference arm CSDM
\begin{alignat}{1}
\mathbf{W}_{11}(\mathbf{r}_{1},\mathbf{r}_{1}) & =\frac{k^{2}\mathbf{T}_{1}^{*}\mathbf{J}_{\mathrm{eff}}\mathbf{T}_{1}^{\mathrm{T}}}{4\pi^{2}d^{2}}\iint_{-\infty}^{\infty}\mathrm{d}^{2}r_{1}^{\prime}\mathrm{d}^{2}r_{2}^{\prime}\gamma(\mathbf{r}_{2}^{\prime}-\mathbf{r}_{1}^{\prime})\nonumber \\
 & \times\exp\left\{ \frac{ik}{2d}\left[\mathbf{r}_{2}^{\prime2}-\mathbf{r}_{1}^{\prime2}-2\left(\mathbf{r}_{2}^{\prime}-\mathbf{r}_{1}^{\prime}\right)\cdot\mathbf{r}_{1}\right]\right\} .\label{eq:w02p11}
\end{alignat}
 With the change of variables $\Delta\mathbf{r}^{\prime}=\mathbf{r}_{2}^{\prime}-\mathbf{r}_{1}^{\prime}$
and $\mathbf{R}^{\prime}=(\mathbf{r}_{1}^{\prime}+\mathbf{r}_{2}^{\prime})/2$,
the integration with respect to $\mathbf{R}^{\prime}$ becomes proportional
to the delta function $\delta(k\Delta\mathbf{r}^{\prime}/d)$.  After
scaling the delta function, integrating over it and using the property
$\gamma(\mathbf{0})=1$, we have   
\begin{alignat}{1}
\mathbf{W}_{11}(\mathbf{r}_{1},\mathbf{r}_{1}) & =\mathbf{T}_{1}^{*}\mathbf{J}_{\mathrm{eff}}\mathbf{T}_{1}^{\mathrm{T}}=J_{0,ii}\mathbf{T}^{i}\label{eq:w04p11}
\end{alignat}
 with $i\in\{x,y\}$. The latter form in equation (\ref{eq:w04p11}) follows
from   $\mathbf{J}_{\mathrm{eff}}=\mathbf{T}^{i}\mathbf{J}_{0}\mathbf{T}^{i}=J_{0,ii}\mathbf{T}^{i}$
and $\mathbf{T}_{1}=\mathbf{I}$. 

To compute the CSDM of the test arm ($\mathbf{W}_{22}$) we return
to the delta-correlated source. Using equations (\ref{eq:et01}), (\ref{eq:weff01}),
(\ref{eq:k01}) and (\ref{eq:weff00p}) we obtain (after integration
over $\mathbf{r}_{2}^{\prime}$) 
\begin{alignat}{1}
\mathbf{W}_{22}(\mathbf{r}_{2},\mathbf{r}_{2}) & =\frac{k^{2}}{4\pi^{2}d^{2}}\int_{-\infty}^{\infty}\mathrm{d}^{2}r_{1}^{\prime}\mathbf{T}_{2}^{*}\mathbf{T}^{*}(\mathbf{r}_{1}^{\prime})\mathbf{J}_{\mathrm{eff}}\mathbf{T}^{\mathrm{T}}(\mathbf{r}_{1}^{\prime})\mathbf{T}_{2}^{\mathrm{T}}.\label{eq:w03p22}
\end{alignat}
Again employing $\mathbf{J}_{\mathrm{eff}}=J_{0,ii}\mathbf{T}^{i}$
the  test arm CSDM becomes 
\begin{alignat}{1}
\mathbf{W}_{22}(\mathbf{r}_{2},\mathbf{r}_{2}) & =\frac{k^{2}J_{0,ii}\mathbf{T}^{j}}{4\pi^{2}d^{2}}\int_{-\infty}^{\infty}\mathrm{d}^{2}r_{1}^{\prime}\left|T_{ji}(\mathbf{r}_{1}^{\prime})\right|^{2}\label{eq:w04p22}
\end{alignat}
 with $i,j\in\{x,y\}$, since $\mathbf{T}^{j}\mathbf{T}(\mathbf{r}_{1}^{\prime})\mathbf{T}^{i}=T_{ji}(\mathbf{r}_{1}^{\prime})\mathbf{T}^{ji}$.
 The traces of equations (\ref{eq:w04p11}) and (\ref{eq:w04p22}) are
used to obtain equations (\ref{eq:trw04p11}) and (\ref{eq:trw04p22}).

\subsubsection{Case B: Polarizer in front of the reference arm detector\label{sub:Optical-element-in}}

When the first polarizer is placed in front of the reference arm detector
the matrix $\mathbf{T}_{1}^{*}\mathbf{J}_{0}$ is proportional to
$\mathbf{T}_{1}=\mathbf{T}^{i}$ when $\mathbf{J}_{0}$ is diagonal.
{[}This proportionality is required to find the main imaging result,
equation (\ref{eq:trw04p}).{]} To obtain equal intensities for both field
components of the source, the light needs to be completely unpolarized.
Light which is partially linearly polarized along either the $x$
or $y$ axis is also sufficient but will result in a lower intensity
source for one of the measurements when $i\in\{x,y\}$ is varied.

 Using equation (\ref{eq:w03p22}) and the constraints $\mathbf{J}_{\mathrm{eff}}=\mathbf{J}_{0}$
and $\mathbf{T}_{2}=\mathbf{T}^{j}$, we note that the term $J_{0,ii}|T_{ji}(\mathbf{r}_{1}^{\prime})|^{2}$
in equation (\ref{eq:w04p22}) is replaced by $J_{0,xx}|T_{jx}(\mathbf{r}_{1}^{\prime})|^{2}+J_{0,yy}|T_{jy}(\mathbf{r}_{1}^{\prime})|^{2}.$
Thus the CSDM for the test arm is 
\begin{alignat}{1}
\mathbf{W}_{22}(\mathbf{r}_{2},\mathbf{r}_{2}) & =\frac{k^{2}\mathbf{T}^{j}}{4\pi^{2}d^{2}}\int_{-\infty}^{\infty}\mathrm{d}^{2}r_{1}^{\prime}\nonumber \\
 & \times\left(J_{0,xx}\left|T_{jx}(\mathbf{r}_{1}^{\prime})\right|^{2}+J_{0,yy}\left|T_{jy}(\mathbf{r}_{1}^{\prime})\right|^{2}\right).\label{eq:w04p22b}
\end{alignat}
 Taking the trace of equation (\ref{eq:w04p22b}) results in equation (\ref{eq:trw04p22b}).

\bibliographystyle{apsrev4-1}

\end{document}